\documentclass[sn-nature]{sn-jnl}

\usepackage{graphicx}%
\usepackage{multirow}%
\usepackage{amsmath,amssymb,amsfonts}%
\usepackage{amsthm}%
\usepackage{mathrsfs}%
\usepackage[title]{appendix}%
\usepackage{xcolor}%
\usepackage{textcomp}%
\usepackage{manyfoot}%
\usepackage{booktabs}%
\usepackage{algorithm}%
\usepackage{algorithmicx}%
\usepackage{algpseudocode}%
\usepackage{listings}%

\theoremstyle{thmstyleone}%
\theoremstyle{thmstyletwo}%

\theoremstyle{thmstylethree}%

\begin{document}

\title[Social media polarization reflects shifting political alliances in Pakistan]{Social media polarization reflects shifting political alliances in Pakistan}


\author[1]{\fnm{Anees} \sur{Baqir}}
\equalcont{These authors contributed equally to this work.}

\author[1]{\fnm{Alessandro} \sur{Galeazzi}}
\equalcont{These authors contributed equally to this work.}

\author[1]{\fnm{Andrea} \sur{Drocco}}

\author*[1,2]{\fnm{Fabiana} \sur{Zollo}}\email{fabiana.zollo@unive.it}

\affil*[1]{\orgname{Ca' Foscari University of Venice}, \country{Italy}}
\affil[2]{\orgname{The New Institute Centre for Environmental Humanities}, \country{Italy}}



\abstract{The rise of ideological divides in public discourse has received considerable attention in recent years. However, much of this research has been concentrated on Western democratic nations, leaving other regions largely unexplored. Here, we delve into the political landscape of Pakistan, a nation marked by intricate political dynamics and persistent turbulence. Spanning from 2018 to 2022, our analysis of Twitter data allows us to capture pivotal shifts and developments in Pakistan's political arena. By examining interactions and content generated by politicians affiliated with major political parties, we reveal a consistent and active presence of politicians on Twitter, with opposition parties exhibiting particularly robust engagement. We explore the alignment of party audiences, highlighting a notable convergence among opposition factions over time. Our analysis also uncovers significant shifts in political affiliations, including the transition of politicians to the opposition alliance. Quantitatively, we assess evolving interaction patterns, showcasing the prevalence of homophilic connections while identifying a growing interconnection among audiences of opposition parties. Our study, by accurately reflecting shifts in the political landscape, underscores the reliability of our methodology and social media data as a valuable tool for monitoring political polarization and providing a nuanced understanding of macro-level trends and individual-level transformations.}




\maketitle

\section{Introduction}

In recent years, there has been a growing concern over the ideological divides that have emerged within our societies in public debates. Researchers, politicians, and public figures have raised alarms about the widening ideological disagreements between opposing factions on various fronts and the potential consequences for public discourse~\citep{falkenberg2022growing, bavel2020using,barbera2015birds,flamino2023political,hare2014polarization,del2017mapping, stella2018bots}. One notable example is the Capitol Hill assault, which occurred during a period when U.S. politics had reached its highest level of polarization since the Civil War~\citep{axelrod2021preventing}. Polarization, in its various forms, has garnered significant attention from scholars across multiple disciplines, resulting in a vast body of literature on the subject~\citep{bavel2020using,iyengar2019origins}.

In particular, political polarization has been extensively studied as a key aspect of societal ideological divides~\citep{druckman2021affective,flamino2023political,bovet2019influence, conover2011political}. Social and political scientists have investigated political divisions for decades, employing various methodologies such as roll call voting records, surveys, or combining these data with other sources~\citep{hare2014polarization, prior2013media}. The emergence of the Internet and social media has  enabled researchers to investigate polarization through digital data, quantifying the divides and tracking their evolution within online communities~\citep{conover2011political,bovet2019influence,stella2018bots, barbera2015tweeting,falkenberg2022growing}. These studies have examined various aspects of social media polarization, including the roles of politicians, news outlets, public figures, and community structures~\citep{barbera2015tweeting,flamino2023political}.
However, most studies have primarily focused on a limited number of countries, primarily Europe and the United States, with little attention given to other geographical areas~\citep{morales2015measuring,najafi2022secim2023,takikawa2017political,recuero2020hyperpartisanship,evkoski2023content}.
Notably, South Asia and the Middle East remain understudied in this context. Existing works have mainly revolved around the ``Arab Spring''~\citep{weber2013secular, wolfsfeld2013social, bruns2013arab, howard2011opening} and have largely overlooked other relevant issues. 

Here, we aim to address this gap by conducting an analysis of political polarization in South Asia, with a specific focus on Pakistan. Since its inception in 1947, Pakistan's political landscape has been characterized by persistent turbulence. Despite being a federal parliamentary democratic republic, military rule has consistently exerted influence over the nation, fostering skepticism towards democratic institutions~\citep{akhtar2022struggle,jan2021rule}. This military involvement stands as a pivotal driver of political polarization in Pakistan, yet it is not the sole contributing factor. In addition, ethno-linguistic groups within the country actively seek recognition, resulting in regional political conflicts that pose substantial challenges to national unity efforts~\citep{shafqat2020pakistan}. Moreover, the presence of religious extremism, exemplified by groups like the Taliban, exacerbates polarization. Extremist actors conduct attacks not only within Pakistan but also across its borders with India and Afghanistan~\citep{javed2022heterogeneous}. In this charged environment, the Pakistani media, instead of promoting critical news assessment, often exhibit pronounced political biases~\citep{akhtar2022struggle}.

Pakistan was ranked 107th in the Democracy Index 2022, with an overall score of 4.13, notably lower than scores exceeding 8, which are typical of fully democratic countries~\citep{economist_democracy}. It is important to note that the concept of political polarization is closely linked to the political system in which it is assessed.  Indeed, polarization in democratic countries can differ significantly from that in flawed democracies, hybrid regimes, or authoritarian states. Therefore, gaining a comprehensive understanding of how political polarization has emerged and evolved over time in Pakistan holds significant importance from multiple perspectives.

In this work, we examine the evolution of political polarization in Pakistan using Twitter data. Between 2018 and 2022, Pakistani politics witnessed significant shifts, marked by the rise of  Pakistan Tehreek-e-Insaf (PTI) -- in English, the ``Pakistan Movement for Justice"-- to power after securing a majority in the 2018 General Elections. Imran Khan led a coalition government with various allies. The formation of the Pakistan Democratic Movement (PDM) in September 2020, an opposition alliance featuring major players like Pakistan Muslim League - Nawaz (PML-N) and the Pakistan People's Party (PPP), added a new dimension to opposition politics. As PTI's allies and some members defected to the PDM, PTI lost its parliamentary majority. Ultimately, this culminated in the ousting of PTI's government in 2022 following a vote of no confidence. This led to the formation of a coalition government under Shehbaz Sharif of PML-N in April 2022. 

To explore these developments, we collected the timelines of members of parliament belonging to the three major political parties from 2018 to 2022. Additionally, we gathered data on the retweets they received to reconstruct interaction networks and measure political polarization. Our study includes a temporal analysis to track the evolution of polarization and the volume of interaction across parties.

Our results highlight a close correspondence between the parties' distances in the latent ideological space and the evolution of Pakistani politics. This correspondence reveals a convergence over time among accounts and audiences of opposition parties that eventually formed a coalition in 2022.
Thus, we prove the parallelism between our ideology estimation in the latent space obtained by social media data and the evolution of parties' political views.
Our research provides valuable insights into the state and evolution of political polarization within a South Asian nation, an area that has received limited attention in the literature. Furthermore, it offers further evidence of the reliability of social media data for tracking the evolution of political polarization, demonstrating a close alignment between temporal analysis and unfolding political events. Finally, this study underscores the adaptability of our methodology, making it applicable to a wide range of scenarios.

\section{Results}
\label{sec:results}
\begin{figure}[ht!]
    \centering 
    \includegraphics[width=\textwidth]{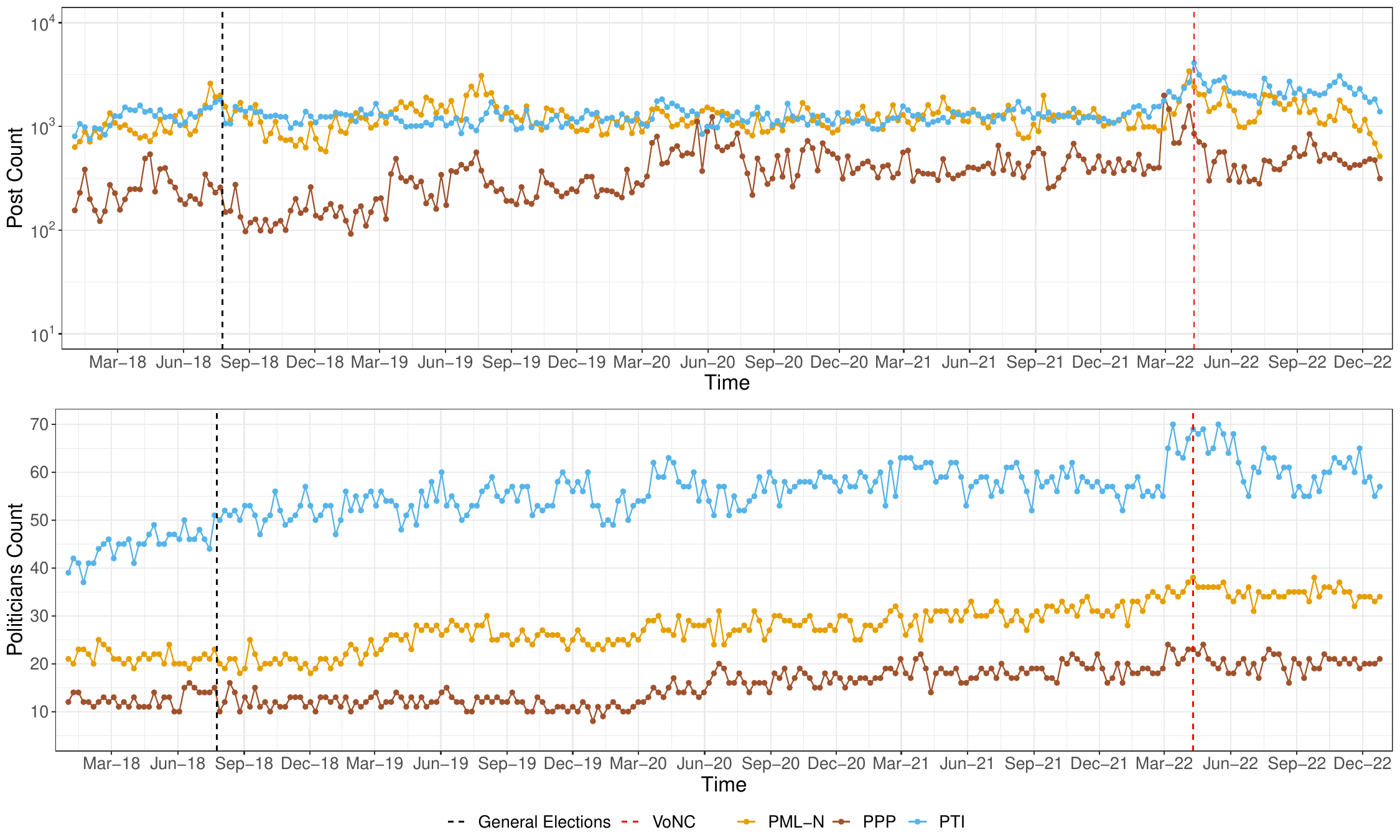}
    \caption{\textbf{Politicians' activity over time.} Evolution of weekly content volume and active accounts. 
    The upper panel shows the weekly number of posts (tweets and retweets) made by politicians of PML-N (yellow), PPP (brown), and PTI (blue). The lower panel reports the number of active accounts belonging to politicians from 2018-2022.}
    \label{fig:post_count}
\end{figure}

We begin our analysis of the Pakistani political landscape on Twitter by depicting the weekly post volume and the number of active accounts per week for members of parliament from the three major political parties: PML-N (yellow), PPP (brown), and PTI (blue), as shown in Figure~\ref{fig:post_count}. Traditionally, the PML-N platform has leaned towards conservatism, advocating principles such as free markets, deregulation, lower taxes, and private ownership. However, in recent years, the party's political ideology and platform have shifted noticeably, becoming more liberal on social and cultural issues. The PPP, on the other hand, has its roots in socialist principles and aims to transform Pakistan into a social-democratic nation. It remains dedicated to promoting equality, pursuing social justice, and maintaining a strong military presence. PTI has declared its primary focus on shaping Pakistan into a model welfare state with principles of Islamic socialism. It is committed to dismantling religious discrimination against Pakistani minorities and identifies itself as an anti-status quo movement advocating Islamic democracy rooted in egalitarianism. However, while in power, PTI has faced criticism for its repression of opposition parties, attempts to curb freedom of speech, and efforts to increase control over the media.

We observe that both PML-N and PTI maintained consistently high, stable, and comparable volumes of tweets over time. 
In contrast, PPP's tweet activity exhibited a gradual increase over time, with notable spikes occurring during June-July 2020 and again in March-April 2022. During these spikes, PPP's tweet volume reached levels comparable to the other two parties. Furthermore, when examining the number of weekly active accounts, we notice a slight upward trend for all three parties. Despite having a significantly smaller number of accounts, PML-N managed to produce a volume of content on par with that of PTI, indicating more intensive social media usage. Similarly, while PPP had fewer accounts, their weekly post count occasionally matched that of the other parties. This suggests that the two opposition parties, PML-N and PPP, made more extensive use of Twitter than PTI, which was the ruling party at the time. Crucially, Figure~\ref{fig:post_count} provides an insightful overview of politicians' Twitter usage, highlighting their consistent presence on the platform. This underscores Twitter's reliability as a valuable data source for studying the evolution of online political discourse.

\subsection{Interactions among parties over time}
\begin{figure}[h]
    \centering
    \includegraphics[width=\textwidth]{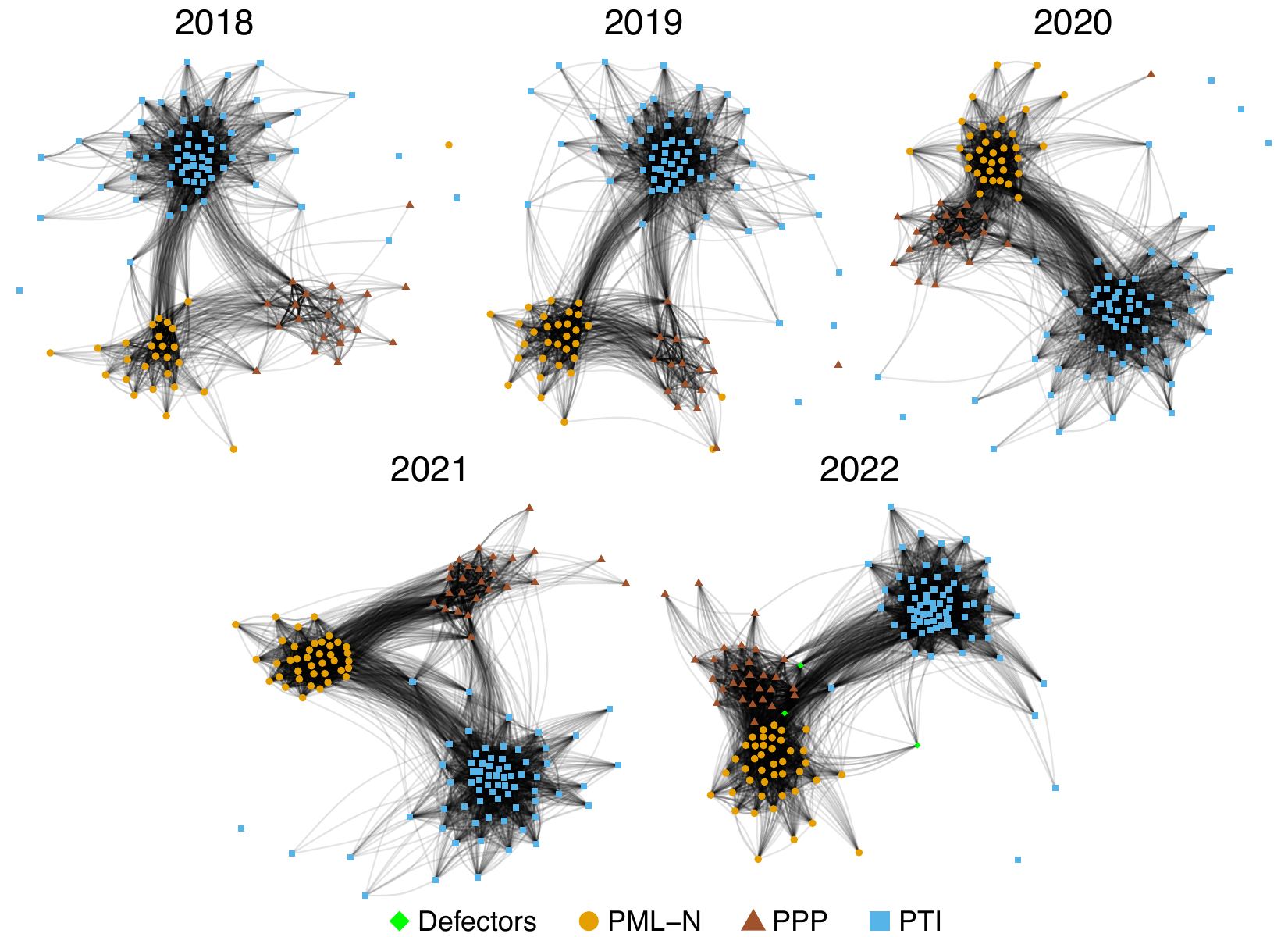}
    \caption{\textbf{Politicians' Audiences Similarity Network.}
    Networks of politicians based on the similarity of their retweeters from 2018 to 2022. Each node is color-coded based on the political party to which it belongs. Each node is color-coded according to the political party to which it belongs. The thickness of the edges represents the strength of the cosine similarity between the retweeter sets of the nodes. To emphasize the most significant connections, we have excluded edges with weights below the median of network weights (see SI for the complete networks). The convergence between PML-N and PPP clusters indicates an increase in intra-party links from 2019 onwards. By 2022, these clusters have become very close. Notably, some nodes (in green), representing defectors from PTI who left the party in 2022 to join the opposition parties, have increased their proximity to the PML-N and PPP clusters while distancing themselves from PTI.}
    \label{fig:networka}
\end{figure}

A commonly employed approach to assess the state of online discourse involves analyzing users' consumption patterns.
Thus, we visualize the divisions within the Pakistani political landscape and and how they have evolved over time by leveraging the similarities among the retweeters of members of parliament (MPs).
We built undirected networks of MP accounts based on their retweeters to unveil similarities in their audiences and then compared these patterns with their political affiliations (for details, see section~\ref{sec:methods}). We generated one network for each year spanning from 2018 to 2022, as shown in Figure~\ref{fig:networka}.  This allowed us to examine the evolution of political divisions through users' consumption patterns.

The results reveal the presence of three distinct clusters, each corresponding to one of the top three parties, indicating a high degree of audience similarity among politicians within the same affiliation. However, it is noteworthy that in 2018, which marked the year of the general elections, the PML-N (yellow dots) and PPP (red triangles) clusters had limited mutual connections. Yet, over time, the connectivity between these two clusters has steadily increased, reaching its peak in 2022. Conversely, the PTI cluster (blue squares) exhibits a lower degree of connectivity with the other two parties, and this connectivity does not appear to increase over time.

These trends are visually evident in the decreasing distance between the PPP and PML-N clusters from 2018 to 2022, while the PTI cluster consistently maintains its distance from the others. These findings imply that an increasing proportion of supporters of both the PML-N and PPP parties have progressively started consuming political content from both parties. However, this pattern does not hold for nodes affiliated with the PTI party, which appear to have a group of supporters primarily or exclusively engaging with the content they produce.

It is noteworthy that some nodes from the PTI cluster, highlighted in green, defected from the party in 2022 and joined the other two parties. This shift in affiliations is reflected in their decreased distance from the PML-N and PPP clusters while substantially increasing their distance from the PTI cluster. The fact that our results accurately reflect the defection of these two politicians confirms the validity of our methodology for analyzing the political discussion. Overall, Figure~\ref{fig:networka} provides an overview of the initial conditions, evolving relationships, and changes in user consumption patterns in response to shifts in political affiliations within the Pakistani political landscape.

\begin{figure}[h]
    \centering
    \includegraphics[width=\textwidth]{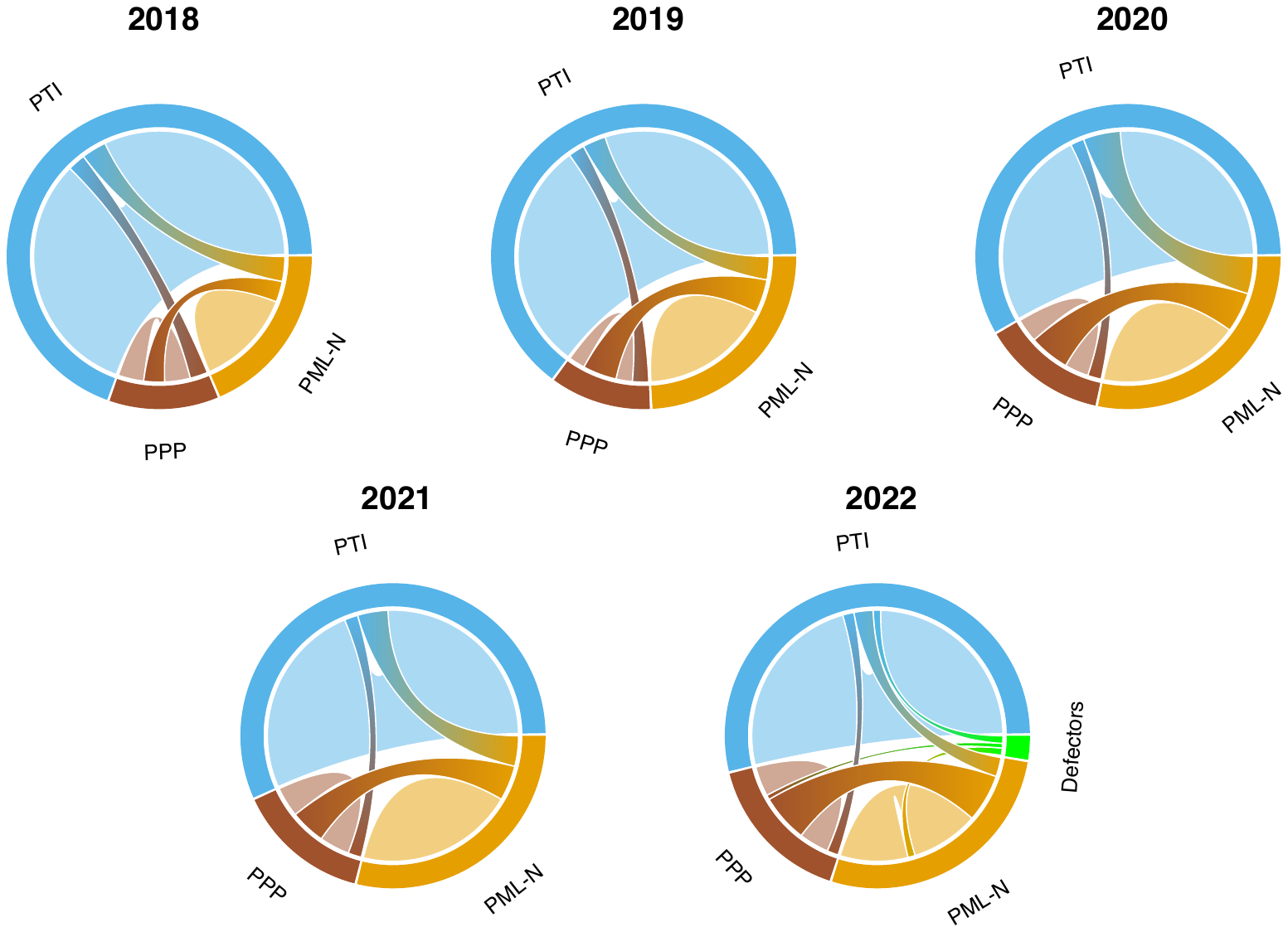}
    \caption{\textbf{Evolution of the shares of connections among political parties.}
     The flows represent the number of network edges in each network, grouped by nodes' political affiliation for each year. While connections within the same party (homophilic connections) are prevalent, there is a noticeable increase in the fraction of connections between the PPP and PML-N parties. This trend aligns with their eventual formation of a coalition.}
    \label{fig:polarization_pol_analysis}
\end{figure}

To quantify the shifts in audience similarity among nodes affiliated with different parties, we conducted an analysis of the changing proportion of connections between opposing factions within the networks depicted in Figure~\ref{fig:networka}. Figure~\ref{fig:polarization_pol_analysis} illustrates the progression of connections between political parties from 2018 to 2022.

Throughout all years, interactions are predominantly characterized by homophilic connections. Notably, links between nodes affiliated with PTI constitute the majority of total network interactions in all years, ranging from 64.2\% in 2018 to 49.5\% in 2022. These PTI connections also represent the highest proportion of links within PTI (over 79\%). Similarly, interactions within the PML-N party primarily consist of connections among members of the same party, accounting for over 47\% of total interactions. Conversely, starting in 2019, PPP accounts have a higher share of links with PML-N nodes (over 44\%) than among themselves (less than 39\%).

Notably, the mutual connections between PPP and PML-N nodes have increased over time, rising from 5.39\% in 2018 to 12.54\% in 2022. Additionally, when considering the share of links relative to each party, the fraction of PPP links connecting to a PML-N node was approximately 32\% in 2018 and increased to nearly 51\% in 2022. Similarly, for PML-N links to PPP, the share rose from 22\% in 2018 to 34\% in 2022.

On the contrary, connections between PTI and other parties exhibit a declining trend, decreasing from 4.60\% to 2.98\% with PPP and from 6.46\% to 4.14\% with PML-N. Despite defections from PTI, defector nodes still share approximately 39\% of their links with PTI, about 24\% with PPP, and roughly 37\% with PML-N (for detailed values, see SI).

To validate our findings, we also analyzed networks constructed from direct retweets among politicians. While the number of nodes and edges is considerably lower in these networks, the analysis produced qualitatively identical results (see SI).
By quantifying the increasing intertwining of PPP and PML-N audiences, our results highlight the convergence of these two parties in the online political discourse, which corresponds with unfolding political events.

\subsection{Quantifying Polarization}
To assess the evolution of the political divisions in the online debate over time, we employed the latent ideology technique, which has proven to be a robust methodology for quantifying users' opinions on various topics~\citep{barbera2015birds,bovet2019influence,flamino2023political,falkenberg2022growing,baqir2023beyond}.
The latent ideology method is designed to infer users' opinions through correspondence analysis. It operates by transforming a weighted bipartite matrix representing interactions between a set of prominent accounts, known as influencers, and other users, projecting them into a one-dimensional real space. Users who interact with similar sets of influencers will be placed close to each other, while those interacting with disjoint sets of influencers will be placed in opposite positions (further details can be found in Section~\ref{sec:methods}). 
This approach enabled us to infer the political ideology of users based on the politicians they retweeted. 
We considered accounts belonging to politicians and users who had retweeted at least three different politicians. We applied the latent ideology algorithm to the weighted bipartite matrix of retweets between users and politicians, obtaining the estimated users' ideologies. 
Furthermore, we calculated the ideological positions of politicians as the median of their retweeters' positions.
To capture the dynamics of the political discourse, we conducted the ideology analysis for each year within the observation period.

\begin{figure}[ht!]
    \centering
    \includegraphics[width=0.7\textwidth]{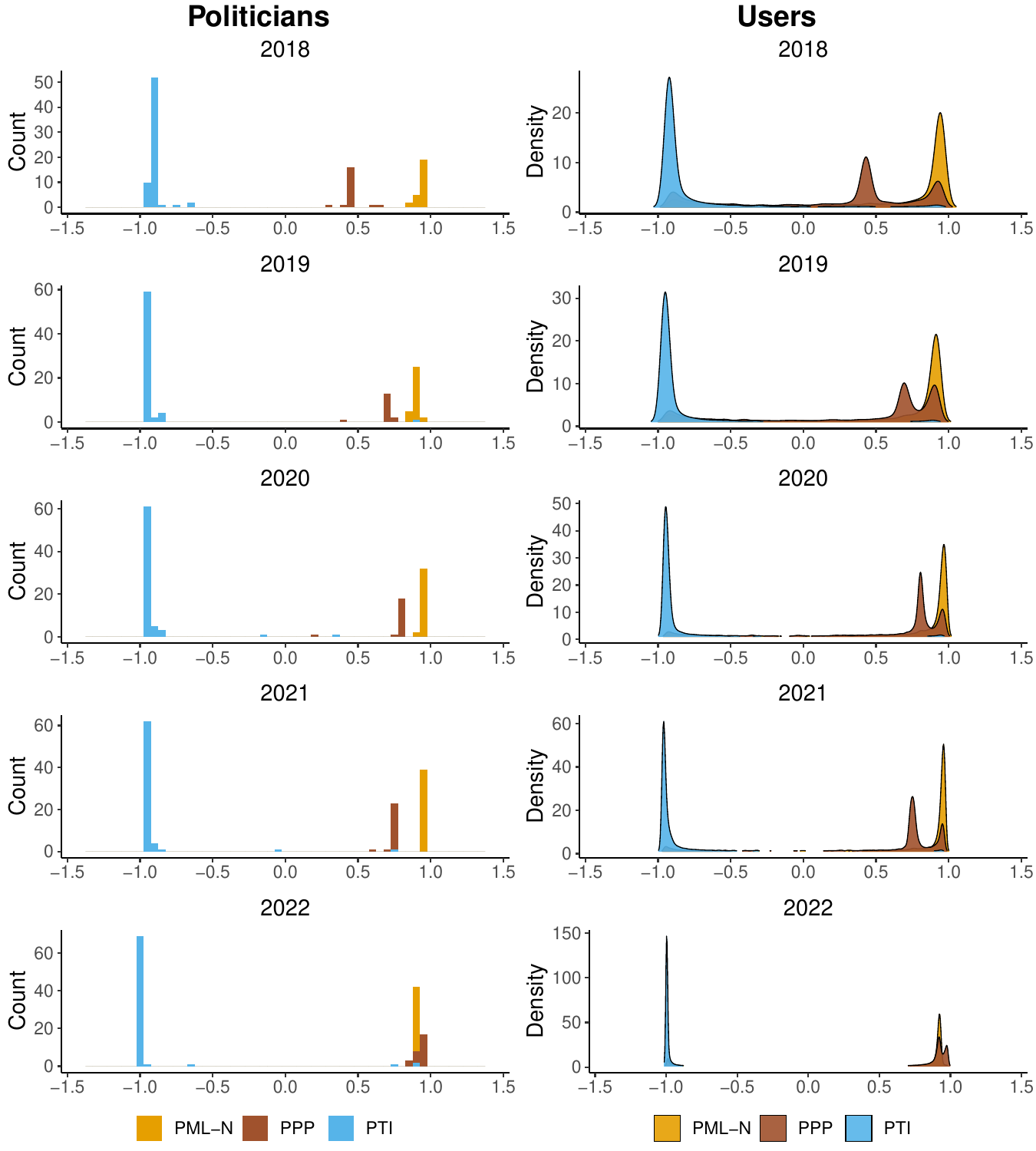}
    \caption{\textbf{Evolution of Politicians' and Users' Latent Ideology.}\\
    Latent ideology estimation for politicians (left column) and users (right column) of the three political parties, based on retweet data from 2018 to 2022. Bars (left panel) represent the count of politicians belonging to one of the three political parties, while curves(right panel) represent the distribution density of their retweeters. Notably, the PML-N (yellow) and PPP (brown) bars are positioned on the right side, indicating a higher audience similarity with respect to PTI. On the left side, the PTI bar (blue) represents a distinct political ideology. Notice the presence of blue points on the right part of the spectrum in 2022, corresponding to the PTI defectors.}
    \label{fig:polarization_pol}
\end{figure}

The results of latent ideology estimation are presented in Figure~\ref{fig:polarization_pol}. On the left side, we present the ideological positions of politicians, while on the right side, we display the distribution of ideologies among users. This distinction arises from the fact that the numbers of politicians affiliated with the three parties are of similar magnitudes, but there is a significant disparity in user counts among the parties. Our focus regarding politicians is to monitor their positions, even at the individual level, especially in cases of defectors. Conversely, when it comes to users, our primary concern is tracking changes in their distribution across the ideological spectrum. Therefore, for users, the priority is unveiling where the audience is most concentrated for each party, rather than merely comparing absolute values among the parties. To aid in this distinction, we color-coded the politicians' affiliations and the ideology distribution of their retweeters. This allows us to track the evolution of both politicians and users' positions within each party.

We notice that polarization dominates the Pakistani political landscape. However, the temporal analysis also reveals an interesting pattern. In 2018, politicians' ideologies display a distinct separation between the three parties, with PTI politicians on one side and two closer yet distinct peaks for PPP and PML-N representatives on the other. 
Users' ideology distribution also exhibits three distinct peaks, each corresponding to one of the three parties.

Notably, users tend to occupy a broader range of the ideological spectrum compared to politicians. Parties' audiences extend into the territory of other parties' peaks, with PPP even having a smaller secondary peak in the same position as the PML-N peak. This observation underscores the close relationship between PPP and PML-N and the presence of a shared audience between these two parties.

Since we assigned users to parties based on the politicians they retweeted, users who retweeted accounts from different parties are counted multiple times. Thus, individuals who primarily retweeted content from one party but occasionally engaged with others contribute to the density of each retweeted party. The presence of PPP users under the PML-N peak indicates shared retweeters and helps explain why PPP and PML-N are closer in ideological space than PTI across all the observation periods.

When we examine the temporal evolution of polarization, we observe that both politicians and users from PPP and PML-N moved closer to each other over time, to the extent that they merged into a single peak in 2022. Meanwhile, the gap between PTI and PML-N remained consistent throughout the observation period.

This is also reflected in the level of polarization, which we measured using Hartigan's dip test (as described in Section~\ref{sec:methods}). The polarization level among politicians increased from 0.14 in 2018 to 0.23 in 2022 (p values $< 0.001$).



Notably, while PPP and PML-N increased their audience similarity and reduced their distance in the latent ideological space over time, in 2022 PTI experienced a dramatic increase in its audience, reflected in the highest differences in its users' density with respect to the other parties. 

\begin{figure}[ht!]
    \centering
    \includegraphics[width=\textwidth]{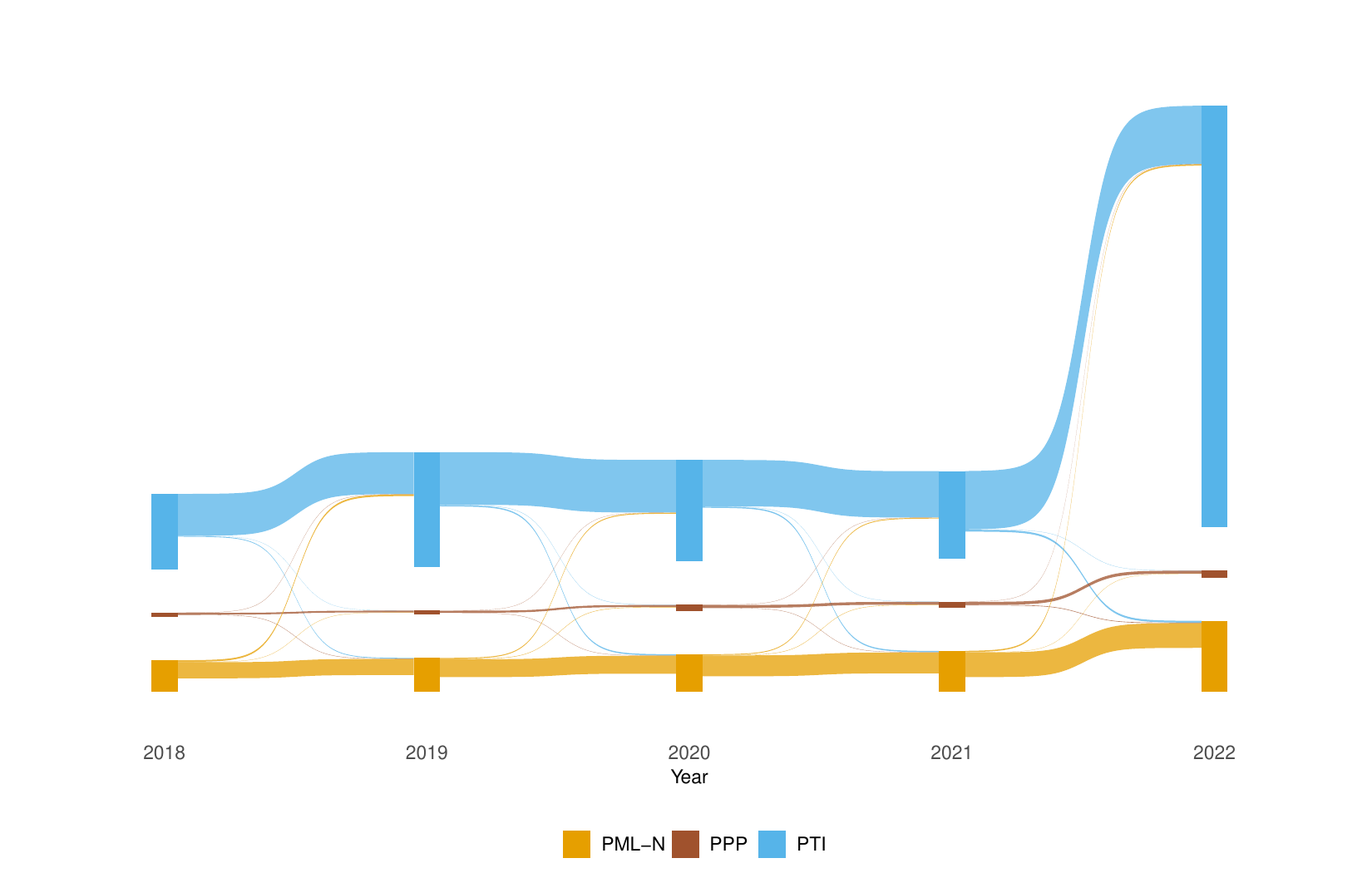}
    \caption{\textbf{Users' political affiliation over time.} Flows widths are proportional to audience changes across political parties over time. 
    The size of each bar represents the proportion of unique accounts associated with a specific political party. The shifts indicate the percentage of users who have changed their political affiliation.}
    \label{fig:leaning_flow}
\end{figure}

To shed light on users' growth over time, we examined the number of users participating in the debate for each political party. We assigned a political affiliation to each user based on the dominant retweeted party. Hence, if a user retweeted accounts from different parties, they were assigned to the party they retweeted the most.
We repeated this procedure for each year to uncover any changes in users' affiliations.
The results are shown in Figure~\ref{fig:leaning_flow}, where for each party the height of each bar corresponds to the number of retweeters, and the flow widths represent shifts in users' political affiliations over time.

Our analysis reveals that users' affiliations have remained relatively consistent over the years, with only minimal shifts from one party to another. Moreover, the number of retweeters for PML-N and PPP experienced a gradual increase over time, particularly in 2022 when they saw more pronounced growth(+92.62\% and +35.78\% respectively). The PTI audience witnessed a notable increase in users in 2019 (+55.71\%), a slight decrease in 2020 (-12.41\%) and 2021 (-10.51\%), but a dramatic surge in 2022(+525.36\%). The reasons behind this substantial growth can be diverse, ranging from a more heated political debate to the increased presence of PTI supporters and automated accounts. While identifying the exact causes is outside the scope of this study, future research in this direction is certainly warranted.

When comparing these results with the evolution of polarization shown in Figure \ref{fig:polarization_pol}, it becomes evident that, while users, in general, have not significantly changed their primary party affiliation, the volume of accounts retweeting content from both PML-N and PPP has steadily increased over time. Conversely, the PTI audience has not shown a significant increase in cross-ideological interactions but has experienced a substantial growth in retweeting exclusively its content.

\section{Discussion}
In this work, we studied the evolution of polarization within the Pakistani political debate on Twitter. We focused on the content generated by accounts affiliated with politicians from the three major Pakistani parties and the interactions they fostered. Our findings highlight the consistent use of Twitter as a communication medium by Pakistani politicians throughout the entire observation period, with their presence on the platform growing over time. Notably, our comparison of account and content production volumes revealed a more active presence from opposition parties (PML-N and PPP) compared to the party in government (PTI).

Moreover, our analysis delved into the similarities among the audiences of different parties and their evolution over time, revealing a convergence between the two opposition parties (PML-N and PPP). Our techniques proved sensitive enough to detect the shift of two PTI representatives who left the party to join the opposition alliance.

Furthermore, we quantified the evolution of interactions between parties, showing that homophilic connections dominated the entire observation period. However, we observed an increasing entanglement among audiences of opposition parties, marked by a growing share of common consumers over time. Consequently, the distance between these two opposition parties, as measured by latent ideology estimation, decreased over time. In contrast, opposition parties maintained a consistent gap in relation to the government party (PTI).

Remarkably, the shift of defectors was clearly discerned through latent ideology estimation, reinforcing its reliability as a tool for measuring polarization in online debates. 

Lastly, we examined shifts in audiences among parties over time, revealing the stability of most users' political affiliations while also emphasizing a surge in PTI retweeters in 2022. 

Our results indicate that, although the majority of users remained loyal to their chosen political party over time, opposition parties experienced an increase in the number of common retweeters. Thus, while users predominantly consumed content from one party, the number of PML-N/PPP users consuming content from both opposition parties increased over time.

The significance of our work extends on multiple fronts, emphasizing its importance in the following three key aspects.
First, we study the intricate evolution of polarization within the Pakistani political discourse on Twitter. While polarization on social media, particularly Twitter, has garnered extensive attention, certain regions like South Asia have remained conspicuously under-explored. Our study contributes to bridging this critical gap, providing valuable insights into the state and dynamic evolution of online polarization in response to unfolding political events. This is especially pertinent in understanding the unique dynamics of polarization in the South Asian context and within a peculiar political environment.

Secondly, our research reinforces the credibility of Twitter data as a robust tool for studying political polarization. It underscores the possibility to not only detect macro-level shifts in political alliances but also capture nuanced individual-level changes in accounts' political affiliations. This adds another layer of validation to the use of Twitter data in political polarization analysis.

Lastly, our work demonstrates the scalability of polarization analysis using social media data across diverse environments.
Recent studies have applied similar techniques to various debates, spanning from politics ~\citep{barbera2015birds,flamino2023political}, to climate change~\citep{falkenberg2022growing}. Despite the cultural and social disparities inherent in these debates, our results affirm the adaptability of such analytical approaches to different contexts. This highlights the utility of the applied methodology in understanding multifaceted issues across various domains.

Naturally, we recognize certain limitations. First, it is important to acknowledge that Twitter may not perfectly mirror real-life circumstances and may not represent a fully representative cross-section of the Pakistani population. Furthermore, our methodology may inadvertently exclude users and politicians who did not generate a substantial volume of data. 

Despite these limitations, we maintain that our results offer a robust and reliable portrayal of the state and evolution of online political polarization in Pakistan, as evidenced by their alignment with real-life events. Moreover, the significance of our work may extend beyond the specific context of Pakistani politics. High levels of polarization, whether in politics, climate change discussions, or vaccine debates, can pose significant threats to public society, potentially resulting in inaction against critical issues or even public unrest. The capability to analyze polarization trends nearly in real-time using social media data holds immense value and can become a tool for informing decision-making to prevent critical situations. Hence, we believe that the availability of social media data for such analysis is of paramount importance, serving both academic inquiries and societal concerns.

\section{Materials and Methods}
\label{sec:methods}

\subsection{Twitter Data}
\label{subsec:data}
In this study, we exploited a dataset consisting of tweets from the Twitter timeline of members of the Pakistani Parliament.
We created a list of active Twitter accounts belonging to members of the Pakistani Parliament by manually gathering their Twitter usernames.
Using the Twitter API for academic research purposes, we collected data from a total of 160 politicians representing the three major political parties in Pakistan, as determined by the number of elected members in the 2018 General Elections~\footnote{\hyperlink{https://ecp.gov.pk/storage/files/1/National\%20Assembly1.pdf}{https://ecp.gov.pk/storage/files/1/National\%20Assembly1.pdf}}. 
The data collection spanned from January 2018 to December 2022, and we exclusively gathered publicly available content from public accounts. To allow the study of polarization and interaction structure, we also collected all retweets for all the tweets whose retweets count is greater than 3.
The detailed breakdown of the dataset is provided in Table \ref{tab:dataset}.

\begin{table}[h]
    \centering
    \resizebox{\textwidth}{!}{%
    \begin{tabular}{c|c|c|c|c|c|c|c}
    \hline \hline
     \textbf{Party} & & \textbf{2018} & \textbf{2019} & \textbf{2020} & \textbf{2021} & \textbf{2022} & \textbf{Total (unique)} \\ \hline
     \multirow{4}{*}{PML-N} & Retweeters & 118,357 & 112,174 & 120,096 & 142,725 & 279,312 & 542,352 \\ 
     & Politicians & 25 & 31 & 34 & 38 & 42 & 44 \\
     & Tweets & 10,456 & 14,046 & 15,393 & 19,739 & 23,767 & 83,401 \\ 
     & Retweets & 2,883,688 & 3,691,028 & 3,782,702 & 4,241,579 & 6,882,825 & 21,481,822 \\ \hline
     \multirow{4}{*}{PPP} & Retweeters & 39,253 & 47,645 & 46,385 & 37,832 & 73,916 & 170,737 \\ 
     & Politicians & 20 & 16 & 21 & 26 & 28 & 34 \\ 
     & Tweets & 3,353 & 3,680 & 7,094 & 6,295 & 8,444 & 28,866 \\ 
     & Retweets & 433,930 & 456,010 & 507,036 & 421,237 & 970,812 & 2,789,025\\ \hline
     \multirow{4}{*}{PTI} & Retweeters & 300,263 & 391,068 & 371,595 & 375,356 & 939,669 & 1,691,124 \\ 
     & Politicians & 66 & 68 & 71 & 70 & 74 & 82 \\ 
     & Tweets & 18,831 & 19,431 & 25,536 & 22,959 & 43,943 & 130,700 \\ 
     & Retweets & 4,469,750 & 6,302,333 & 6,181,935 & 5,499,317 & 48,576,191 & 71,029,526 \\ 
    \hline \hline
    \end{tabular}
    }
    \caption{Breakdown of the dataset by Party and year.}
    \label{tab:dataset}
\end{table}

\subsection{Network Construction}
\label{subsec:network}

We constructed interaction networks among politicians based on retweets information. 
We focused solely on retweets, since unlike quote tweets or comments that may not express support for the original tweet's content, retweets are usually utilized and considered a form of endorsement~\citep{falkenberg2022growing,bovet2019influence,flamino2023political,cinelli2021echo}.
For each year, we built an undirected weighted graph $G$, in which nodes represent politicians belonging to one of the three parties, and edges the retweeters' similarity among them. Using one year of data, we started by creating a matrix $R_{y}$ for each year, with retweeters as rows and politicians as a column, whereas $y \in$ \textit{\{2018, 2019, 2020, 2021, 2022\}}. The entry $r_{i,j}$ of $R_{y}$ is the number of times user $i$ retweeted content posted by politician $j$ in the year $y$. We then computed the cosine similarity for each pair of columns to quantify the retweeters' similarity for each pair of politicians. Thus, the weight $w_{a,b}$ of the edge between node $a$ and $b$ in the graph $G$ is equal to:

\[w_{a,b} = \frac{r_a \cdot r_b}{\Vert r_a \Vert \Vert r_b \Vert}\]

where $r_a$ and $r_b$ are the two column vectors of politicians $i$ and $j$, respectively. It should be noted that $w_{a,b}$ $\in$ [0, 1] since all the matrix entries are non-negative.

Finally, we excluded all the 0-degree nodes and deleted all the edges with a weight below the median of all edge weights. This approach enabled us to capture the strongest similarities among politicians' audiences across the years.

\subsection{Latent Ideology Estimation}
\label{subsec:lie}
 
To estimate the ideological stance of users in the debate, we start from the latent ideology algorithm proposed in~\citep{barbera2015birds, barbera2015tweeting}. Following the studies already conducted in this field~\citep{falkenberg2022growing, flamino2023political}, we consider retweets instead of follower/following relationships as interaction since retweets have been found to be good indicators of content endorsement ~\citep{falkenberg2022growing, flamino2023political}. The latent ideology algorithm requires the selection of a set of accounts, called influencers, which critically affects the ideology estimation results. Since we aim at quantifying the political stance of users, we chose members of parliament accounts as influencers set.  On these accounts and their retweeters, we applied the Correspondence Analysis algorithm~\citep{greenacre2010correspondence}, which follows three steps: (i) Construction of the interaction matrix $A$, (ii) normalization of the matrix, and (iii) singular value decomposition. 
We constructed a matrix $A$, whose elements $A_{ij}$ represent the number of retweets user $i$ directs toward influencer $j$. Once $A$ is known, we normalize it as follows. First, we divide by the total number of retweets, obtaining:  

\begin{equation}
    P=\frac{A}{\sum_{ij} A_{ij}}. 
\end{equation}

Then, we define the following quantities: 

\begin{equation}
    \begin{cases}
    \textbf{r} = P \textbf{1}, \\
    \textbf{c} = \textbf{1}^T P, \\
    D_r = \text{diag}(\textbf{r}),\\
    D_c = \text{diag}(\textbf{c}),
    \end{cases}
\end{equation}

and we perform the following normalization operation: 

\begin{equation}
    S = D_r^{-1/2}(P- \textbf{r}\textbf{c}) D_c^{-1/2}
\end{equation}

For the third step, we perform a singular value decomposition of the form $S= U \Sigma V^T$, where $U, V$ are orthogonal matrices and $\Sigma$ is a diagonal matrix containing the singular values of $S$. Finally, we take the latent ideology of user $i$ to be the $i$-th entry of the first column of the orthogonal matrix $U$, while the retweeters' median ideology represents the latent ideology of an influencer.

\subsection{Hartigan’s Diptest}
\label{subsec:hd}

Hartigan's dip test serves as a nonparametric examination for assessing the presence of multiple modes in a distribution drawn from a sample ~\citep{hartigan1985dip}. It computes the maximum difference across all sample points between the distribution function that minimizes this difference while remaining unimodal and the actual empirical distribution function. The outcome of the test provides a value denoted as $D$, which measures the extent of multimodality, along with a statistical significance value represented as $P$.



\backmatter

\bmhead{Supplementary information} This article has an accompanying supplementary file. 

\bmhead{Acknowledgments} A.G. and F.Z. acknowledge support from the IRIS Research Coalition (UK government, grant no. SCH-00001-3391).

\bibliography{sn-bibliography}

\section*{Supplementary Information}
\begin{table}[ht!]
    \centering
    \begin{tabular}{c|c|c|c|c|c}
    \hline

    \hline
       \textbf{Parties}  & \textbf{2018} & \textbf{2019} & \textbf{2020} & \textbf{2021} & \textbf{2022} \\ \hline

       PML-N - PML-N & 51.70\% & 53.29\% & 47.73\% & 54.67\% & 49.01\% \\ \hline

       PML-N - PPP & 21.97\% & 27.65\% & 27.12\% & 24.22\% & 34.47\% \\ \hline

       PML-N - PTI & 26.33\% & 19.06\% & 25.15\% & 21.11\% & 11.37\% \\ \hline

       PML-N - Defectors & \multicolumn{4}{c|}{} & 5.15\% \\ \hline

       \multicolumn{6}{c}{} \\ \hline

       PPP - PPP & 39.94\% & 24.70\% & 30.77\% & 38.41\% & 32.02\% \\ \hline

       PPP - PML-N & 32.40\% & 50.86\% & 51.72\% & 44.55\% & 50.92\% \\ \hline

       PPP - PTI & 27.66\% & 24.44\% & 17.51\% & 17.04\% & 12.11\% \\ \hline

       PPP - Defectors & \multicolumn{4}{c|}{} & 4.95\% \\ \hline

       \multicolumn{6}{c}{} \\ \hline

       PTI - PTI & 85.31\% & 85.39\% & 79.71\% & 81.77\% & 84.46\% \\ \hline

       PTI - PML-N & 8.58\% & 8.61\% & 14.87\% & 12.67\% & 7.05\% \\ \hline

       PTI - PPP & 6.11\% & 6.00\% & 5.43\% & 5.56\% & 5.10\% \\ \hline

       PTI - Defectors & \multicolumn{4}{c|}{} & 3.39\% \\ \hline

       \multicolumn{6}{c}{} \\ \hline

       Defectors - Defectors & \multicolumn{4}{c|}{} & 0.44\% \\ \hline

       Defectors - PML-N & \multicolumn{4}{c|}{} & 36.73\% \\ \hline

       Defectors - PPP & \multicolumn{4}{c|}{} & 23.89\% \\ \hline

       Defectors - PTI & \multicolumn{4}{c|}{} & 38.94\% \\ \hline

    \hline
    \hline
    \end{tabular}
    \caption{Percentage of links between Parties}
    \label{tab:links_parties}
\end{table}

\begin{figure}[ht!]
    \centering
    \includegraphics[width=\textwidth]{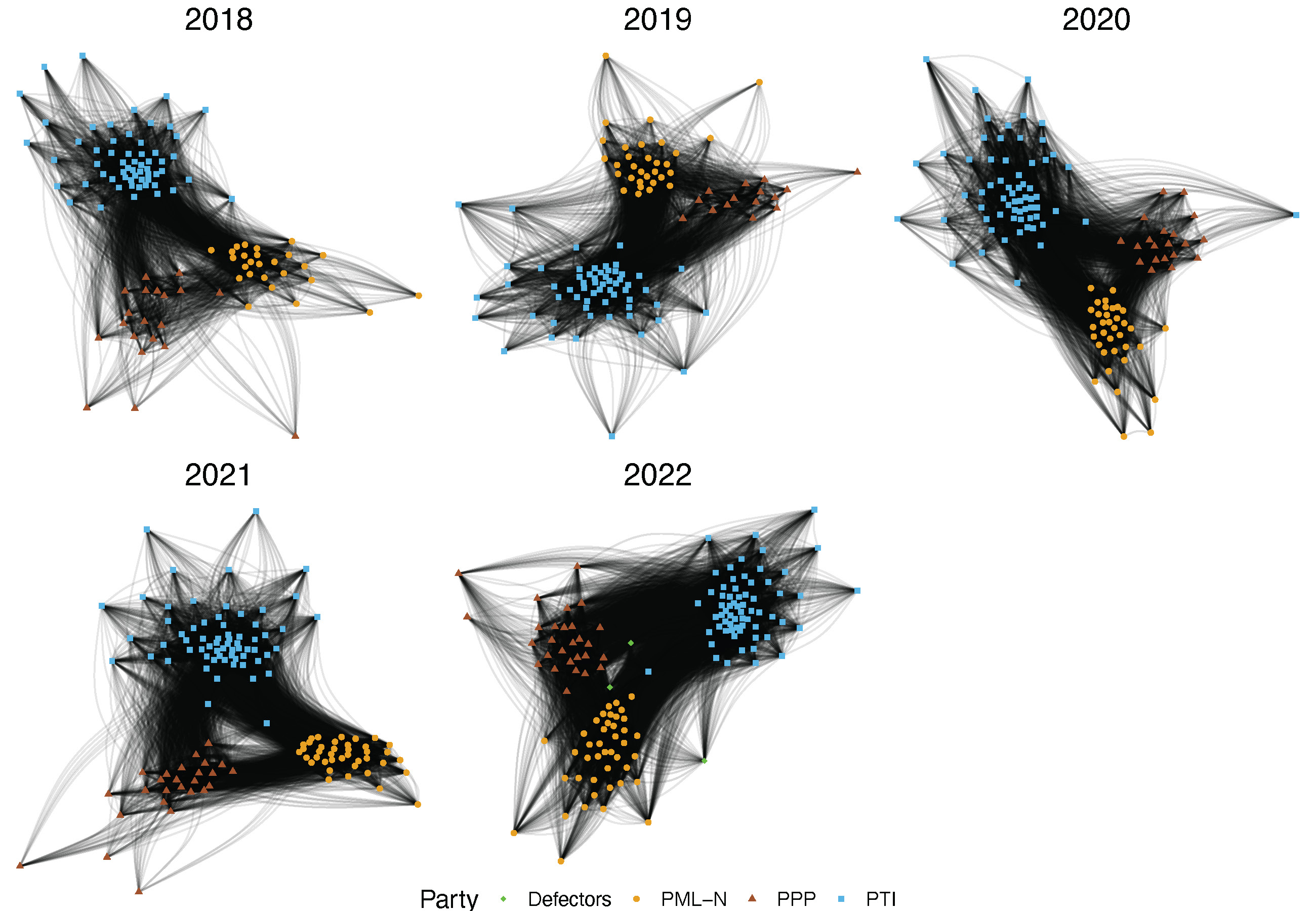}
    \caption{Cosine Similarity networks of politicians’ retweeters from 2018 to 2022 including all edges. Each node is color-coded based on the political party to which it belongs. The thickness of edges between nodes represents the strength of the cosine similarity between the retweeters that propagate the content of politicians. The convergence between PML-N and PPP clusters reveals an increase in intra-party links from 2019 onwards, with them becoming almost a unique cluster in 2022. Throughout the period, certain nodes from PTI consistently appear distant from the core of their cluster compared to those of the other two parties. Noticeably, some nodes (Defectors, in green) from PTI left the party in 2022 and joined the other two parties, reflected in an increased proximity to the PML-N and PPP cluster and an increased distance from PTI.}
    \label{fig:networkb}
\end{figure}

\begin{figure}
    \centering
    \includegraphics[width=\textwidth]{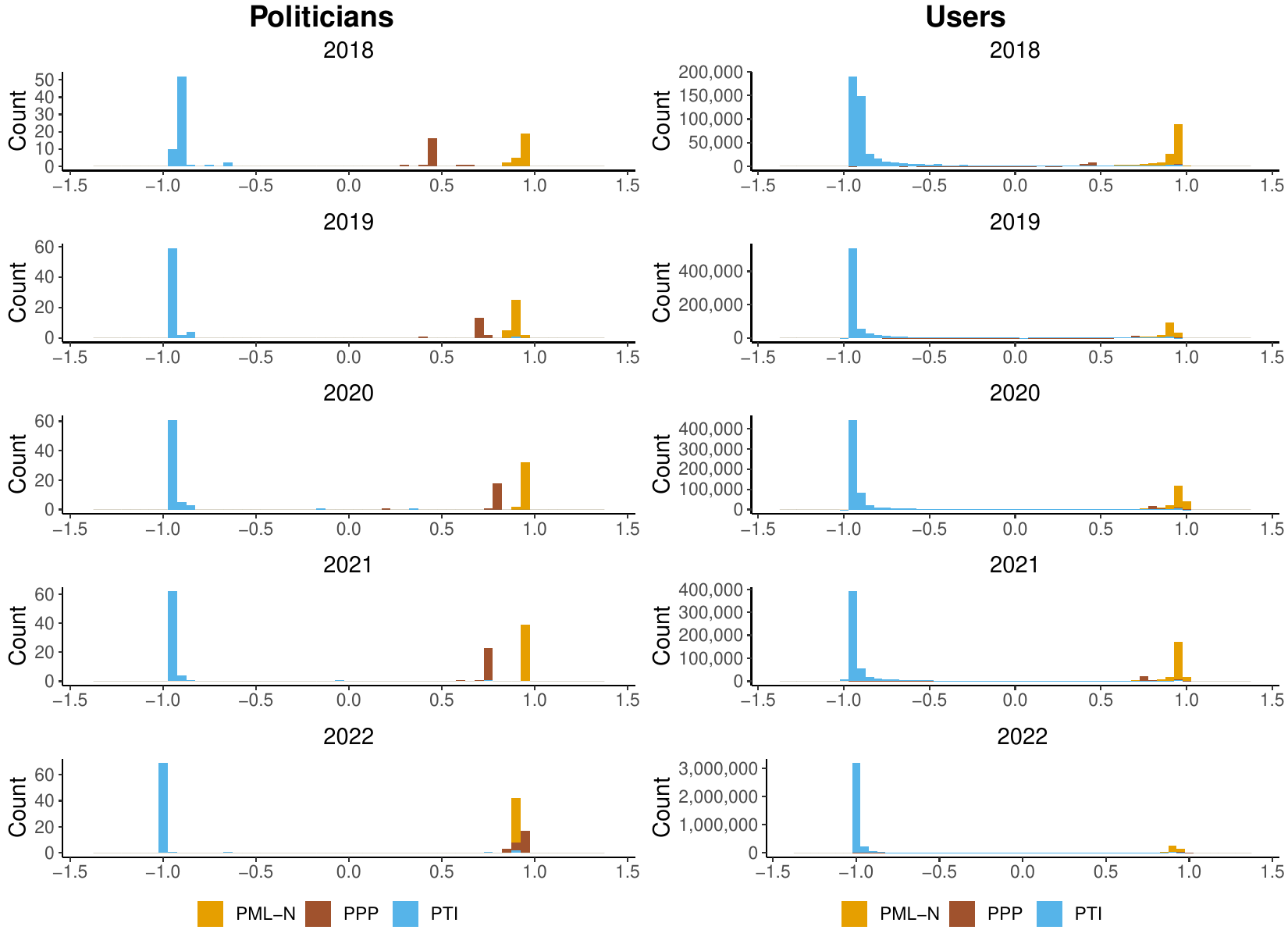}
    \caption{{Evolution of Politicians' and Users' Latent Ideology.}\\
    Latent ideology estimation for politicians (left column) and users (right column) of the three political parties, based on retweet data from 2018 to 2022. Bars (left panel) represent the count of politicians belonging to one of the three political parties, while bars(right panel) represent the count of their retweeters. Notably, the PML-N (yellow) and PPP (brown) bars are positioned on the right side, indicating a higher audience similarity with respect to PTI. On the left side, the PTI bar (blue) represents a distinct political ideology. Notice the presence of blue points on the right part of the spectrum in 2022, corresponding to the PTI defectors.}
    \label{fig:ideology_count_SI}
\end{figure}

\subsection*{Politicians retweets interaction networks}
\begin{figure}[ht!]
    \centering
    \includegraphics[width=\textwidth]{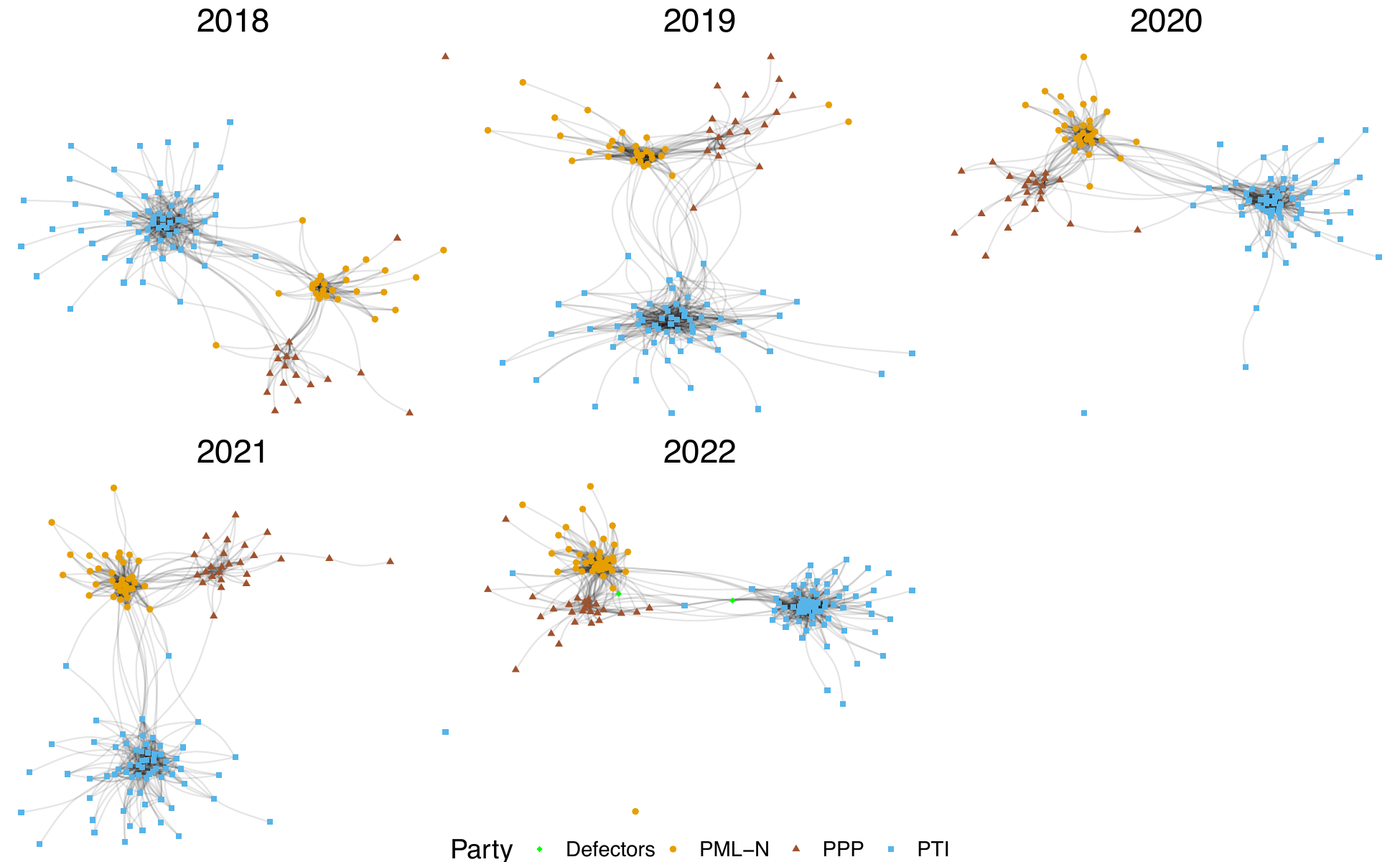}
    \caption{Network of politicians based on retweet interactions from 2018 to 2022. Nodes are color-coded to represent political affiliations: PML-N in yellow, PPP in brown, and PTI in blue. The thickness of edges between nodes indicates the frequency of retweets between politicians, showcasing the level of interaction and endorsement. A noticeable convergence between PML-N and PPP clusters can be observed, signifying an increase in intra-party links from 2019 onwards and leading them to form almost a singular cluster by 2022. Contrarily, some nodes (Defectors, highlighted in green) in 2022 show a distinctive movement towards the PML-N and PPP cluster, indicating a shift from its original affiliation with PTI.}
    \label{fig:networkc}
\end{figure}

\begin{figure}[h]
    \centering
    \includegraphics[width=\textwidth]{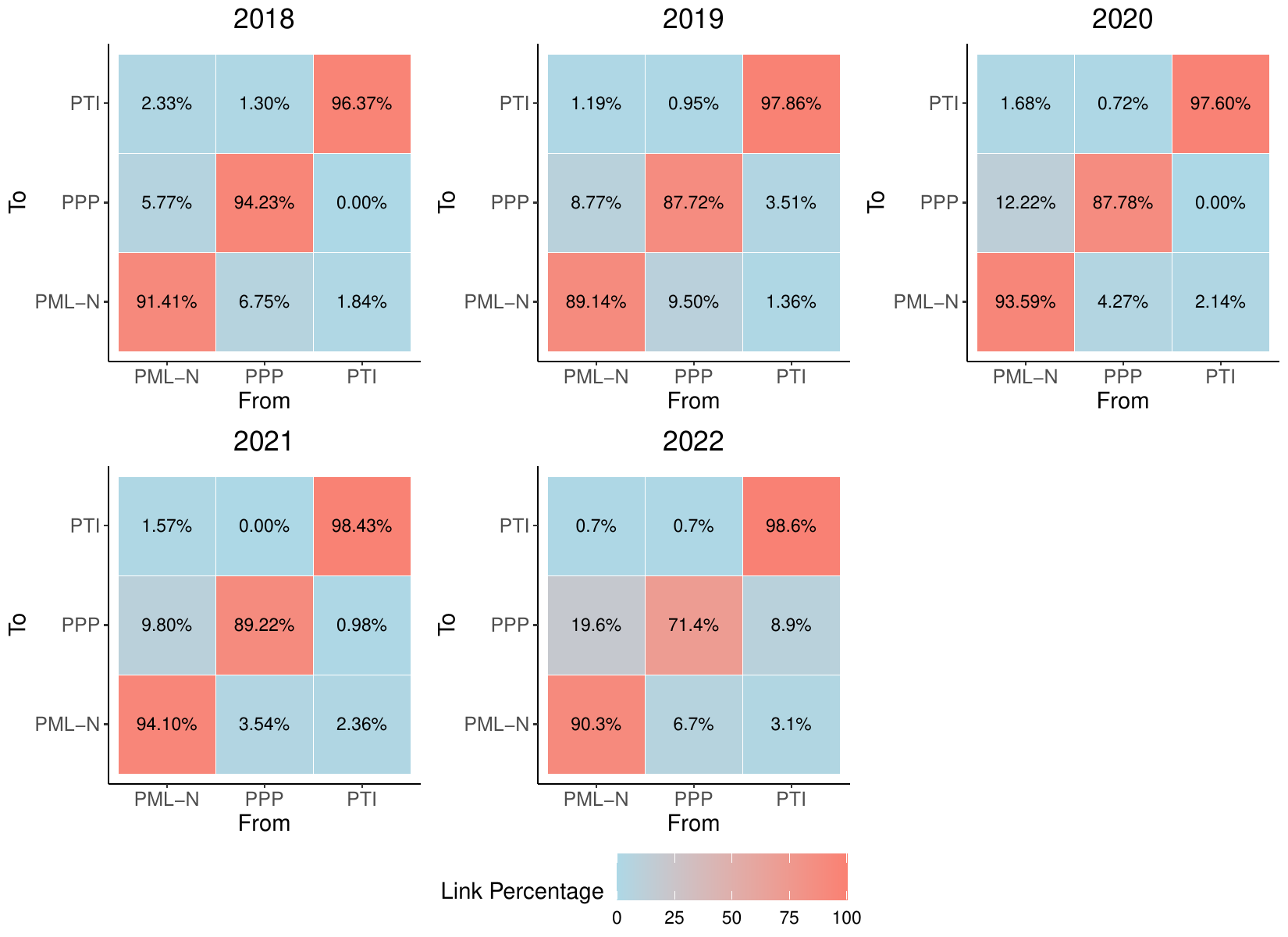}
    \caption{Retweet interaction percentages between political parties from 2018 to 2022. The x-axis (From) represents the origin party, while the y-axis (To) showcases the party being retweeted. Each tile's shade indicates the strength of interaction, derived from the percentage of total retweets. Intra-party retweets remain consistently high for all parties across the years, with PTI showing the highest intra-party interaction in 2022 at 99.28\%. A consistent trend of increasing interaction between PML-N and PPP can be noted, culminating in a significant 19.64\% of PPP's links to PML-N in 2022. In 2022, the 'Defectors' node predominantly retweeted PTI (58.33\%) but also had noteworthy interactions with both PML-N and PPP, at 16.67\% each.}
    \label{fig:politicians_links}
\end{figure}

In the main paper, we illustrated the results of the analysis of politicians' networks built on retweeters' similarity. Here we show a similar analysis of the politicians' networks built on retweet information. Hence, an edge here represents a retweet between two politicians, and edges are weighted according to the number of retweets. Notice that the networks here are directed. Despite a lower volume of data, the results are qualitatively identical to the main paper, showing a progressive entanglement between PPP and PML-N parties and a shift in the PTI defectors, as shown by Figures~\ref{fig:networkc} and~\ref{fig:politicians_links}. 

\end{document}